\documentstyle[preprint,prl,aps]{revtex}
\begin{document}
\draft
\title{Scattering from  Solutions of Star Polymers}
\author{C. M. Marques}
\address{R.P.- C.N.R.S., Complex Fluids Laboratory, Cranbury, NJ 08512-7500, U.S.A.}
\author{T.  Charitat}
\address{Universit\'e Joseph Fourier, L.P.N.S.C, C.N.R.S.\\
25, avenue des Martyrs, B.P. 166, F-38042 Grenoble Cedex 09, France}
\author{D. Izzo}
\address{Instituto de F\'\i sica da Universidade Federal do Rio de Janeiro\\
21945-970 Rio de Janeiro, R.J. Brazil}
\author{E. Mendes}
\address{Laboratoire d'Ultrasons et de Dynamique des Fluides Complexes \\
4, rue Blaise Pascal, 67070 Strasbourg, France}
\date{\today}
\maketitle
\begin{abstract}
We calculate the scattering intensity of dilute and semi-dilute solutions of star
polymers. The star conformation is described by a model introduced by Daoud and
Cotton. In this model, a single star is regarded as a spherical region of a semi-dilute
polymer solution with a local, position dependent screening length. For high enough
concentrations,  the outer sections of the arms overlap and build a semi-dilute
solution (a sea of blobs) where the inner parts of the actual stars are  embedded.  The
scattering function is evaluated following a method introduced by Auvray and de Gennes.
In the dilute regime there are three regions in the scattering function: the 
Guinier region (low wave vectors, $q R \ll 1$) from where  the radius of
the star can be extracted;  the intermediate region ($1 \ll q R \ll f^{2/5}$) that
carries the signature of the form factor of a star with $f$ arms: $I(q) \sim
q^{-10/3}$; and a  high wavevector zone ($q R \gg f^{2/5}$) where the local swollen
structure of the polymers gives rise  to the usual $q^{-5/3}$ decay. In the
semi-dilute regime the different stars  interact strongly, and the scattered intensity
acquires two new features: a liquid peak that develops at a reciprocal position
corresponding to the star-star distances; and a new  large wavevector contribution of
the form $q^{-5/3}$ originating from the sea of blobs. 
\end{abstract} 

\narrowtext 

A star polymer is a branched macromolecule with a small central core from where emerge
several end-attached  linear chains~\cite{gres2}. Not only can star polymers be
regarded as prototypes of branched systems, but also can solutions of star polymers be
considered as  model systems to study polymer colloidal stabilization or the effects
of internal segment density on the behaviour of viscoelastic solutions. Experimental
work in dilute solutions includes studies of   viscoelastic behaviour  and chain
conformation  by rheology\cite{hadj,roov}, light
scattering\cite{hadj,roov,roov1,huber,khas}  and small angle neutron scattering
(SANS)\cite{khas}. The  chain dimensions and structure factors in semidilute
solutions  or melts has also been studied by SANS\cite{will,mend,rich}. Theoretical
work on the conformation of star polymers in good solvents was pioneered by Daoud and
Cotton~\cite{daou} and Birshtein and Zhulina~\cite{birs}.  Renormalization group
calculations~\cite{vlah,miya,free}, Monte Carlo~\cite{rey,bato} and molecular dynamics
simulations~\cite{gres,rich2} corroborate the Daoud-Cotton (D-C) picture at least in
the dilute regime where the simulations have been performed. However, only a limited
amount of information is available to describe the scattering intensity $I({\bf q})$
of star polymers. Analysis of the scattering data in the litterature is performed 
with postulated expressions or limited to the Guinier regime from where the star
radius of gyration can be extracted~\cite{gast}. In this paper we  study the different
factors which contribute to the scattering intensity of a star solution at arbitrary
concentrations.   We follow a method introduced by Auvray and de
Gennes~\cite{auvr,auvr1} to evaluate the intensity scattered by an adsorbed polymer
layer. This method allows for an explicit computation of the scattered intensity of a
star solution  in the full wavevector range $0<q<\left(b(b^{3}/v)\right)^{-1}$, where
$b$ is the Kuhn length and $v$ the excluded volume parameter. 

The structure of a star polymer with $f$ arms of polymerization index $N$ can be
described by the Daoud Cotton~\cite{daou} model. Attachment of
the chains to a central core effectively forces the local polymer density to be
everywhere inside the star above overlapping concentration. The star can therefore be
described as a semi-dilute solution~\cite{dege,witt}, with a local, position dependent
screening length
$\xi(r)$, where $r$ is the distance from the center.  Pictorially, we
represent this by associating with each arm a string of blobs of  increasing size
$\xi(r)$. The radial dependence of the blob size $\xi(r)$ can be
obtained   by noticing that at a distance $r$ from the center there are  $f$ blobs of
cross section $\xi(r)^2$ occupying a total area of $4\pi r^2$. The blob size thus
varies as $\xi(r)\simeq r f^{-1/2}$ and the corresponding local polymer volume fraction
as $\phi_s(r) \simeq f^{2/3}(b/r)^{4/3}$. Note that there is a crowded region of size
$R_c\simeq b f^{1/2}$ in the middle of the star where the concentration reaches one. 
The size of the star can be obtained from monomer conservation  $N f b^3 = 4
\pi\int_0^R r^2 \ dr \phi_s(r)$. Neglecting the small core region one gets $R\simeq a
N^{3/5} f^{1/5}$. Figure 1 sketches the structure of  the star solution above the
overlapping concentration $\phi^{\star} = f N b^3/R^3$. The inner part of each
individual star is still described by a position dependent correlation length $\xi(r)$
and the corresponding concentration profile $\phi_s(r) \simeq f^{2/3} (b/r)^{4/3}$.
However, the profile doesn't vanish at a distance $R$  but levels off to a constant 
value $\phi_p$ at a distance $R_{sd}$, determined from the condition
$\phi_s(R_{sd})=\phi_p$. The concentration $\phi_p$ is the polymer concentration in
the region known as the sea of blobs, where the outer sections of the arms overlap. The
stars embedded in the sea of blobs occupy a volume fraction $\Phi_s$ and  the sea of
blobs the remaining fraction $1- \Phi_s$. Above the overlapping concentration
$\phi^\star$, the values of the polymer concentration in the sea of blobs $\phi_p$,
of the fraction occupied by the stars $\Phi_s$ and of the radius of the embedded stars
$R_{sd}$ can be calculated from mass conservation:
\begin{eqnarray}\label{phisdef}   
\Phi_s \simeq  \left\lbrack{{\phi^\star \over \phi}}\right\rbrack^{5/4} \qquad 
\phi_p \simeq \phi \qquad {\rm and} \qquad R_{sd} \simeq R \left\lbrack
{{\phi^\star \over \phi}}\right\rbrack^{3/4}.   
\end{eqnarray} 
Notice that  an increase of the polymer concentration $\phi$ leads to an increase
of the fraction of space occupied by the sea of blobs  and a shrinkage of the
inner star dimensions.

The scattering intensity due to the chains in solution is given by 
\begin{eqnarray}\label{intdef} 
I({\bf q})=<a^{*}({\bf q})a({\bf q})>, 
\end{eqnarray} 
where $a({\bf q})$ is the amplitude scattered by the ${\cal N}$
monomers in the system and $< >$ denotes an  average   taken over all possible
configurations. The scattered amplitude is written as
\begin{eqnarray}\label{scaamp}
 a(q)=\sum_{i}^{\cal N}
(n_{i}-n_{s}) \exp\{{j{\bf q}.{\bf r}_{i}}\}
\end{eqnarray}
with $n_{i}-n_{s}$ the difference between the
 scattering lengths of the $i^{th}$-monomer and the solvent. We assume hereafter that
all monomers  have equal scattering lengths, $n_{0}$. Defining   $\phi({\bf r})$ as
the monomer volume fraction at position $\bf r$, expression (\ref{intdef}) can be
written as   
\begin{eqnarray} \label{iofq}
I({\bf q})={c \over {\cal N}} \int d^{3}r \exp\{ {j{\bf q}.{\bf r}}\} \int
d^{3}r'<\rho({\bf r'})\rho({\bf r}+ {\bf r'})>,
\end{eqnarray}
with  $\rho({\bf r})=\left(n_0-n_{s}\right)\phi({\bf r})$. The scattering
amplitude $\rho({\bf r})$ of  the semidilute solution of polymer stars described above
can be represented as
\begin{eqnarray}\label{scaampli}
\rho({\bf r})=\sum_{i=1}^{N_{s}} \delta ({\bf r}-{\bf r}_{i})\otimes
\left[\rho_{s}({\bf r}) \Pi({\bf r}) \right] + \rho_{p}({\bf r})\left[1-
\sum_{i=1}^{N_{s}}\delta ({\bf r}-{\bf r}_{i})\otimes \Pi({\bf r}) \right],
\end{eqnarray}
\noindent 
where $N_s = {\cal N}/(N f)$ is the number of micelles in the volume ${\cal V}$, 
the symbol $\otimes$ denotes the convolution integral $ f({\bf r}) \otimes g({\bf r})
\equiv \int d^{3}r'f({\bf r}-{\bf r}')g({\bf r}')$ and $\Pi({\bf r})$ the step function
 $ \Pi({\bf r}) = 1 \ {\rm for\ } r\leq R \ $ ; $\ \Pi({\bf r})=0  \ {\rm for\ }
r>R$. The first term on the r.h.s. of equation~\ref{scaampli} effectively locates
$N_s$ spherical inner parts of stars at positions ${\bf r}_i$. The second term
accounts for the monomers in the sea of blobs. The scattering amplitudes of the inner
parts of the star  $\rho_{s}({\bf r})$ and of the sea of blobs  $\rho_{p}({\bf r})$ can
be expressed as the sum of an average  contribution and of a fluctuating part
\begin{eqnarray}\label{scafinal} 
\rho_{s(p)}({\bf r})=(n_{0}-n_{s}) \phi_{s(p)}({\bf r})
+(n_{0}-n_{s})\delta \phi_{s(p)}({\bf r})
\end{eqnarray}
By inserting expressions (\ref{scaampli}) and (\ref{scafinal}) into equation
(\ref{iofq}) we  evaluate the total scattering intensity under the following
approximations: i) density fluctuations in different stars are uncorrelated. ii)
density fluctuations in the sea of blobs are uncorrelated to density fluctuations in
the inner part of the stars. This is a reasonable assumption except perhaps for the
correlations between the outer blob layer of the stars and the neighbouring blob layer
in the sea of blobs.  The total scattering intensity can be written as a sum of three
terms  
\begin{eqnarray}\label{ifinalofq}
I({\bf q})=(n_{0}-n_{s})^{2}{\cal V} \left[I_{pp}({\bf q})+I_{ss}({\bf q})+I_{sp}({\bf
q})\right]
\end{eqnarray}
The first  term $I_{pp}$ accounts for density fluctuations in the sea of blobs
\begin{eqnarray}\label{ipp} 
I_{pp}({\bf q})=(1-\Phi_{s}) <\delta\phi_{p}^{2}({\bf q})>,
\end{eqnarray}  
the second term $I_{ss}$ accounts for density fluctuations in the inner part of the
stars
\begin{eqnarray}\label{iss}
I_{ss}({\bf q})=\Phi_{s}<\delta\phi_{s}^{2}({\bf q})>,
\end{eqnarray}
and the third term $I_{sp}$ expresses correlations between  the different stars:
\begin{eqnarray}\label{isp} 
I_{sp}({\bf q})=\Phi_{s}S({\bf q}) \Delta\phi({\bf q})^{2}.
\end{eqnarray}
This last expression contains the stars structure factor  $S({\bf q})= 1/
N_{s} \sum_{m,n=1}^{N_{s}} \exp\{{j{\bf q}({\bf r}_{m}-{\bf r}_{n})}\}$ that
expresses correlations between the positions of the stars in the solution. It
is weighted by the form factor of the inner parts of the stars 
$\Delta\phi({\bf q})^{2}=(\phi_{p}({\bf q})-\phi_{s}({\bf q}))^{2}$. We do not
attempt here a calculation of the structure factor $S({\bf q})$, our results provide
instead a systematic  approach for extracting $S({\bf q})$ from the experimental data. 

The results presented above can also be used  for dilute solutions of polymer stars.
Formally, one sets the sea of blobs contribution equal to zero and takes the limit
$S({\bf q}) \to 1$. The volume fraction occupied by the stars is in this limit 
$\Phi_s = \phi/\phi^\star$. 

We now calculate explicitly the polymer contributions to the three different terms
(\ref{iss})-(\ref{isp}). The contribution from polymer-polymer correlations in  the sea
of blobs can be written as the fourier transform of $g(r)$, the pair correlation
function for density fluctuations. This is a function which decays algebraically as
$r^{-4/3}$ and vanishes rapidly for distances larger than the correlation length $\xi_p
= \xi(\phi_p)$: $g(r) \sim \exp\{-r/(\sqrt{27/20} \xi_p)\}/r^{3/4}$. The numerical
value $\sqrt{20/27}$ is chosen such that the correlation length can be directly 
measured from the lorentzian decay of the structure facture at low wave vector 
$<\delta\phi_{p}^{2}({\bf q})> \sim (1+ (q \xi_p)^2)^{-1}$. Performing the fourier
transform of the correlation function leads to
\begin{eqnarray}\label{delpp} 
<\delta\phi_{p}^{2}({X})> = {\rm Cte} \left\lbrack {\xi_p \over
b}\right\rbrack^{1/3} \sqrt{5\over 3} {\sin \left\{ {2\over 3} \arctan \sqrt{27/20} X
\right\}\over (X^3 + 27/20 X^5)^{1/3}} 
\end{eqnarray}
with the dimesionless wavevector $X = q \xi_p$.  For small wavevectors ($X\ll 1$) one
has by construction $<\delta\phi_{p}^{2}({X})>\simeq {\rm Cte} (\xi_p/b)^{1/3} (1
+X^2)^{-1}$ while for large wavevectors ($X\gg1$) one gets
$<\delta\phi_{p}^{2}({X})>\simeq {\rm Cte} \sqrt{5/4}\  (20/27)^{1/3} (\xi_p/b)^{-4/3}
q^{-5/3}$.

The contribution from the density fluctuations inside the stars can be regarded as the
scattering from a semidilute solution with a position dependent correlation length:
\begin{eqnarray}\label{delss} 
<\delta\phi_{s}^{2}({X})> = {\rm Cte} \left\lbrack {\xi_p \over
b}\right\rbrack^{1/3} \sqrt{5\over 3} 3 \int_\alpha^1 y^{7/3} dy {\sin \left\{ {2\over
3} \arctan \sqrt{27/20} X y \right\}\over ((X y)^3 + 27/20 (X y)^5)^{1/3}} 
\end{eqnarray}
with  $\alpha = R_c/R_{sd}$, the ratio   between the compact core radius $R_c$ and the
inner star radius $R_{sd}$. The form of this contribution is similar to the contribution
from the sea of blobs. For instance, in the large wavevector limit ($X \gg 1$) one has
$<\delta\phi_{s}^{2}({X})> =  9/5 <\delta\phi_{p}^{2}({X})>$.

The form factor $\Delta\phi({\bf q})^{2}$ in equation (\ref{isp}) is calculated from 
\begin{eqnarray}\label{formfactdef}  
\Delta\phi({\bf q})^2 =  {12 \pi\over R_{sd}^3}
\left\lbrack \int_0^{R_{sd}} r^2 dr {\sin q r \over qr} (\phi(r)-\phi_p)
\right\rbrack^2  \end{eqnarray}
with $\phi(r)= 1$ for $r<R_c$ ; $\phi(r)= f^{2/3} (b/r)^{4/3}$ for $R_c<r<R_{sd}$ and
$\phi_p$ a constant over all the integration range. Performing the integral leads to
\begin{eqnarray}\label{formfact} 
\Delta\phi(Z)^2 = {4 \pi R_{sd}^3\over 3}
\left\lbrack {b\over\xi_p}\right\rbrack^{8/3} \left\lbrack\alpha^{5/3} g(\alpha
Z)-g(Z)+ {3\over Z^{5/3}}\int_{\alpha Z}^Z dy   {\sin y \over y^{1/3}} \right\rbrack^2 
\end{eqnarray} 
where we defined a dimensionless wavevector $Z = q R_{sd}$. The asymptotic behaviour
of the form factor for large wavevectors ($Z\gg 1$) is $\Delta\phi(Z)^2\sim 
R_{sd}^{-1/3} \xi_p^{-8/3} q^{-10/3}$. 

Comparison of the three factors described above show that the scattered intensity is
dominated at high $q$ vectors by the $q^{-5/3}$ behaviour that is due, in dilute
star solutions, to concentration fluctuations inside the star and, in   semidilute
star solutions,  to concentration fluctuations in both the inner part of the stars and
in the sea of blobs. At low wave vectors we predict for dilute solutions a Guinier
regime followed by a $q^{-10/3}$ slope in a log-log representation, due to the form
factor of the stars. For semidilute solutions, correlations between different stars are
important, and the form factor multiplies the structure factor which measures those
correlations. In practice this must lead  to the appearence of one or more  correlation
pics in the Guinier zone or above. The crossover between the $q^{-5/3}$ and the
$q^{-10/3}$ regimes is expected to occur at wave vectors $q R_{sd} \sim f^{2/5}$ in 
semidilute solutions and at $q R \sim f^{2/5}$ in dilute solutions.

The general form of the scattering intensity of a polymer star solution, as outlined
above, is in good agreement with results from   both  experiments and  simulations.
In the following paragraph we quantitatively confront our predictions to experimental
data from reference~\cite{mend}. The stars considered in reference~\cite{mend} are
synthesized by the so-called core-first method. In this method,  a core of
divynyilbenzene (DVB) is synthesized and from the surface of the core, anionic
polymerization of polystyrene takes place. This method has the advantage of providing
high functionality objects, but with the inconvenient of a high  polydispersity  of 
the functionality (polydispersity of core masses). Also, since the linear chains are
grown from, sometimes, massive DVB nodules, steric repulsion between monomers in the
early stage of the polymerization are probably responsible for branches which are
longer than the value targeted during synthesys. In the present case, the branch mass
was targeted to be $M_n\sim 2 \times 10^4$.

We plot in figure 2 a typical scattering curve from a semidilute star solution. The
intermediate and large $q$ ranges clearly exhibit slopes $-10/3$ and $-5/3$ as
predicted. In order to extract the structure factor of the stars solution we first
subtract the large wavevector contributions. The unknown constant of equations
(\ref{delpp}) and (\ref{delss})  can be obtained from data on semidilute solutions of
linear polymers. Indeed, the scattering at small angles of linear polymers in a
semi-dilute regime is given by an Orstein-Zernicke equation containing the same
multiplicative constante of equation (\ref{delpp}) and (\ref{delss}) . In our case we
extracted the numerical value of the constant from data on polystyrene in deuterated
toluene~\cite{mend2}. We fixed the ratio
$\alpha$ at the value $0.05$. The  curves obtained after subtraction are presented
in figure 3. For the more diluted sample, $\phi=0.055$, the slope of data in a log-log
representation approaches the theoretical slope of $-10/3$. Deviations from this slope
for the more concentrated samples migth have several origins. Second or higher order
pics in the structure factor would fall in this range, therefore introducing oscillant
deviations from a pure linear slope in the log-log plot. Also, the quality of the stars
may be here of importance, particularly the size,  shape and functionality of the star
cores. A few large cores with a small number of  attached arms would also contribute
significantly to the scattered intensity in the intermediate  range. 

As a final step we divide data from figure 3 by the calculated form factor of
equation (\ref{formfact}). The structure factors obtained by this method are plotted
in figure 4 for five different polymer concentrations. The functions obtained display
the usual features of liquid structure factors, in particular   the  position of the
peak increases with concentration as $\phi^{1/3}$\cite{mend}.  It is also
important to stress that the procedure of subtraction is extremely sensitive to both
the chemical quality of  the stars and the correspondent quality of the scattered
data. We expect that structure factors of better quality can be extracted from data of
a carefully defined chemical nature. Such structure factors could then be confronted to
existing theories of liquid structuration.

In this paper we have calculated the scattering intensity of a  star-polymer solution,
both in the dilute and semidilute regimes. The structure of the solution  and
the conformation of the stars has been described by the Daoud-Cotton~\cite{daou} model,
and the intensity calculated following a method introduced by Auvray and de
Gennes~\cite{auvr,auvr1}. Our results show that the scattering intensity can be
described as the sum of three terms. The first term is the product of the structure
factor of the solution by the form factor of the star. We did not evaluate the structure
factor for a star solution, but our results provide a  procedure that allows its
extraction  from experimental data. The form factor is the fourier transform
of the average concentration profile of the star. This term gives rise to a
caracteristic $q^{-10/3}$ wavevector dependence in the  intermediate $q$ range
($1 \ll qR_{sd} \ll f^{2/5}$). The second and third terms are independent of the
structure factor and measure concentration fluctuations in the inner parts of the
stars and in the sea of blobs. They have both a $q^{-5/3}$ dependence, and dominate
the scattering spectra at high wavevectors $ qR_{sd} \gg f^{2/5}$. Our data is
consistent with available experimental data on star solutions.  

{\sl Acknowledgments} We would like to thank J.F. Joanny for many fruitful discussions.

\begin{figure} [t]
\caption{Schematic representation of a semi-dilute solution of star-shaped polymers.
The average concentration profiles between two neighbouring stars is also represented.}
\end{figure}

\begin{figure} [t]
\caption{Scattering intensity as a function of the scattering vectror in
a log-log  scale for a solution of star polymers. The concentration is
$50$mg/cm$^3$ and the average functionality (overestimated) is $130$. The star
branch mass targeted in synthesis was $M_n = 2\times10^4$.  Straight lines of slopes 
$-5/3$ and $-10/3$ are also represented in the figure.  Data from ref [8].}
\end{figure}

\begin{figure} [t]
\caption{Scattering intensity as a function of the scattering vector in
a log-log  scale for solutions of very polidisperse star polymers at
different concentrations when $q^{-5/3}$ terms
have been subtracted. The slope $-10/3$ is also shown for every data set.}
\end{figure}

\begin{figure} [t]
\caption{Structure factors of solutions of  polydisperse stars at
different concentrations obtained  from the procedure described in the
text.}
\end{figure}

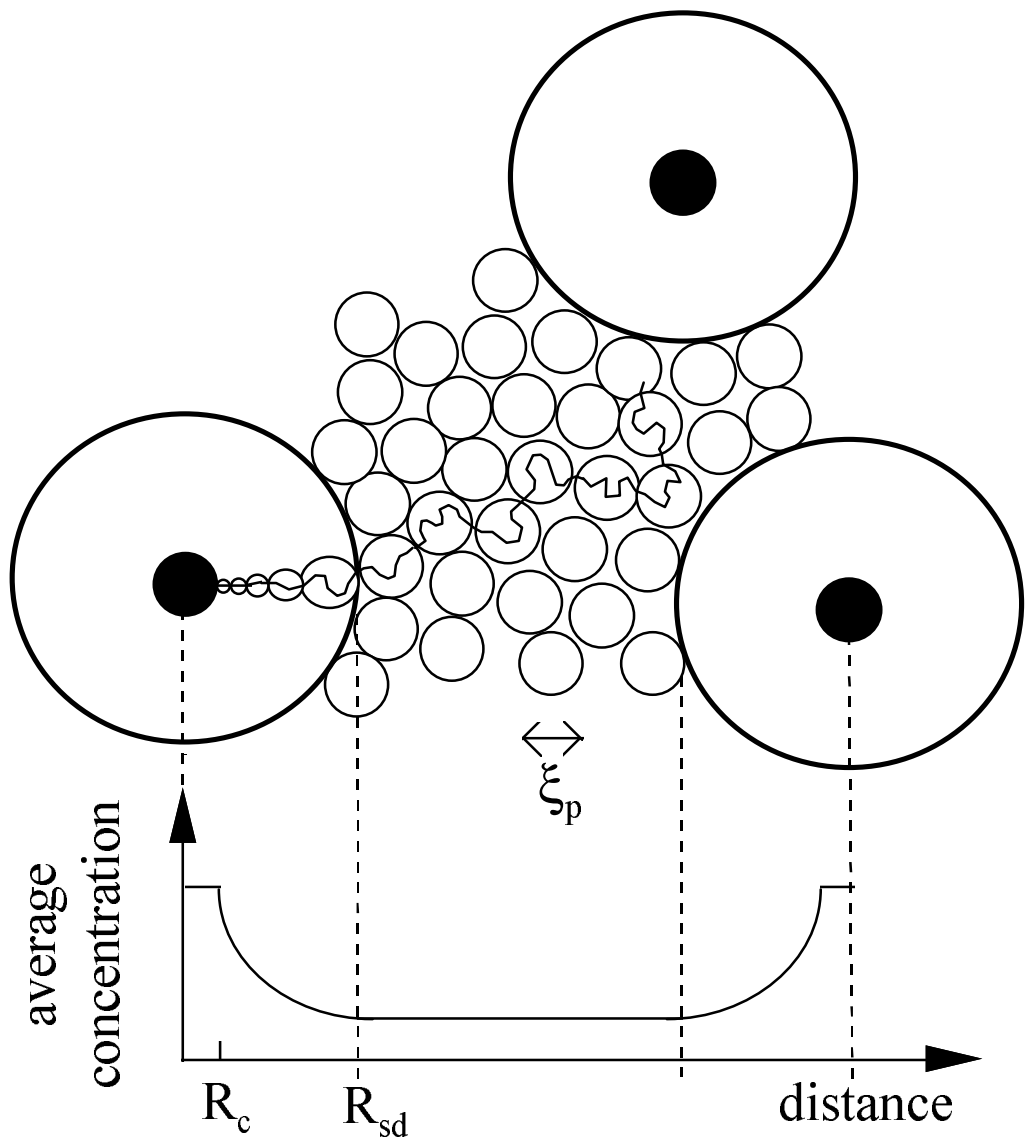
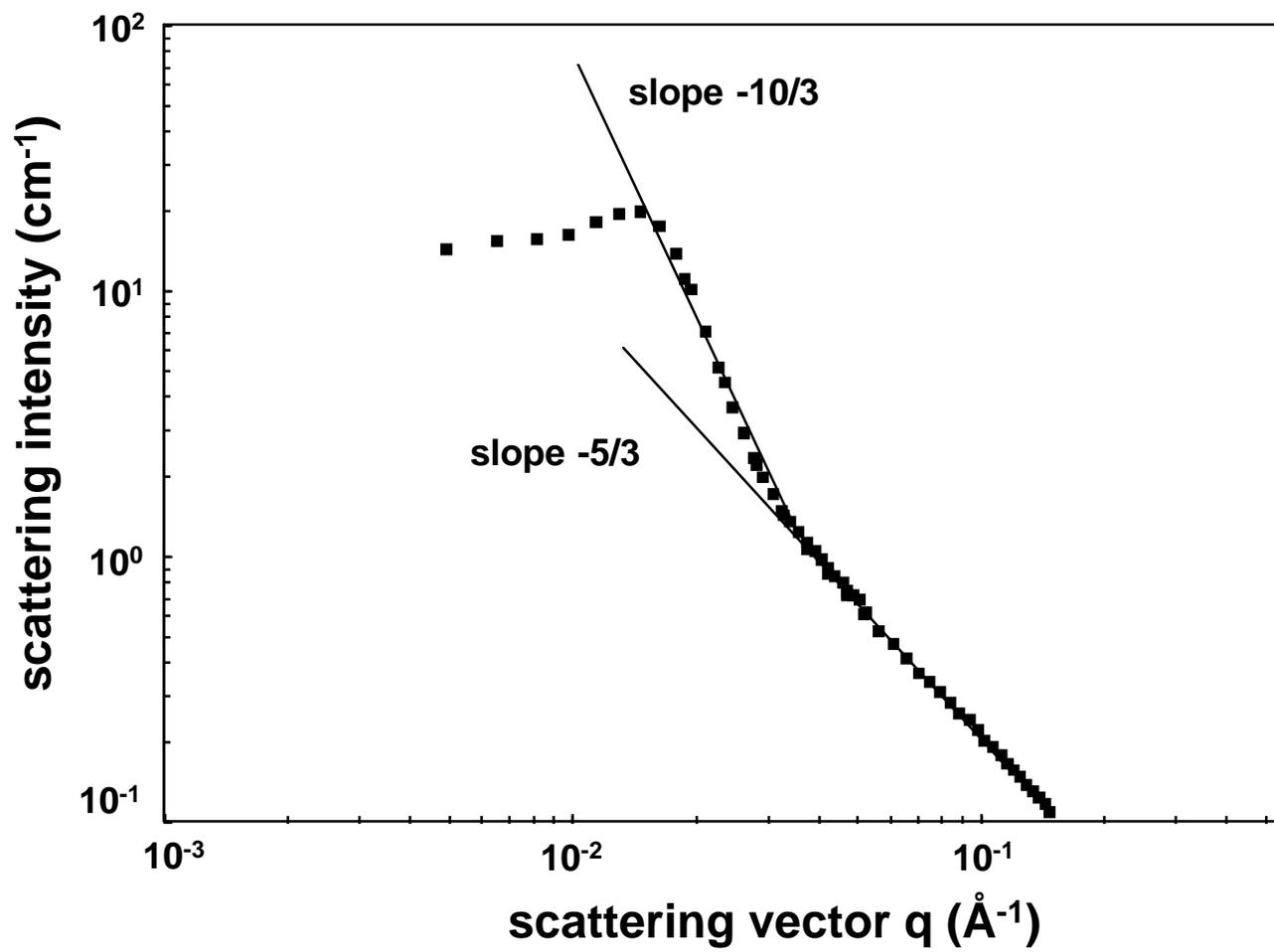
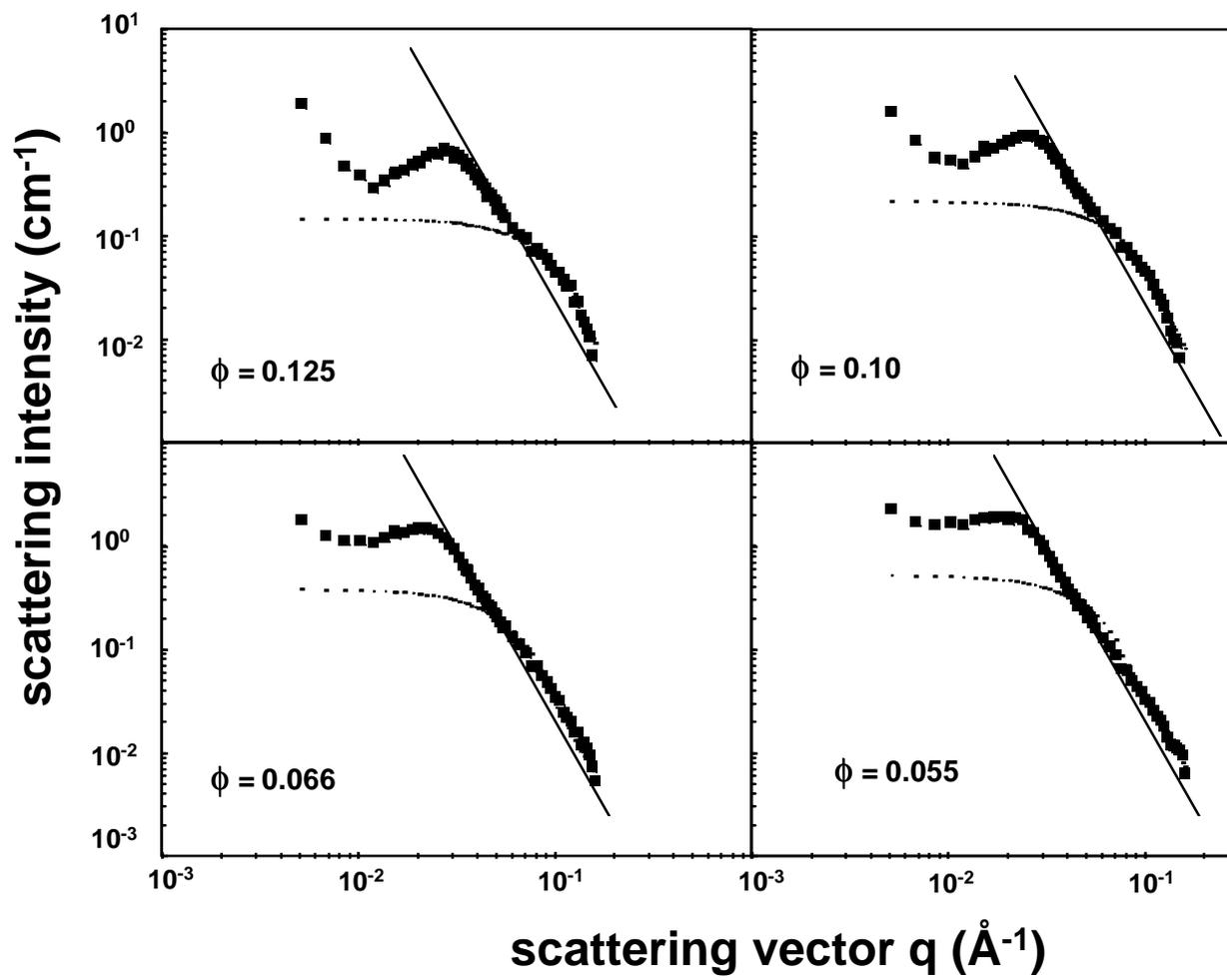
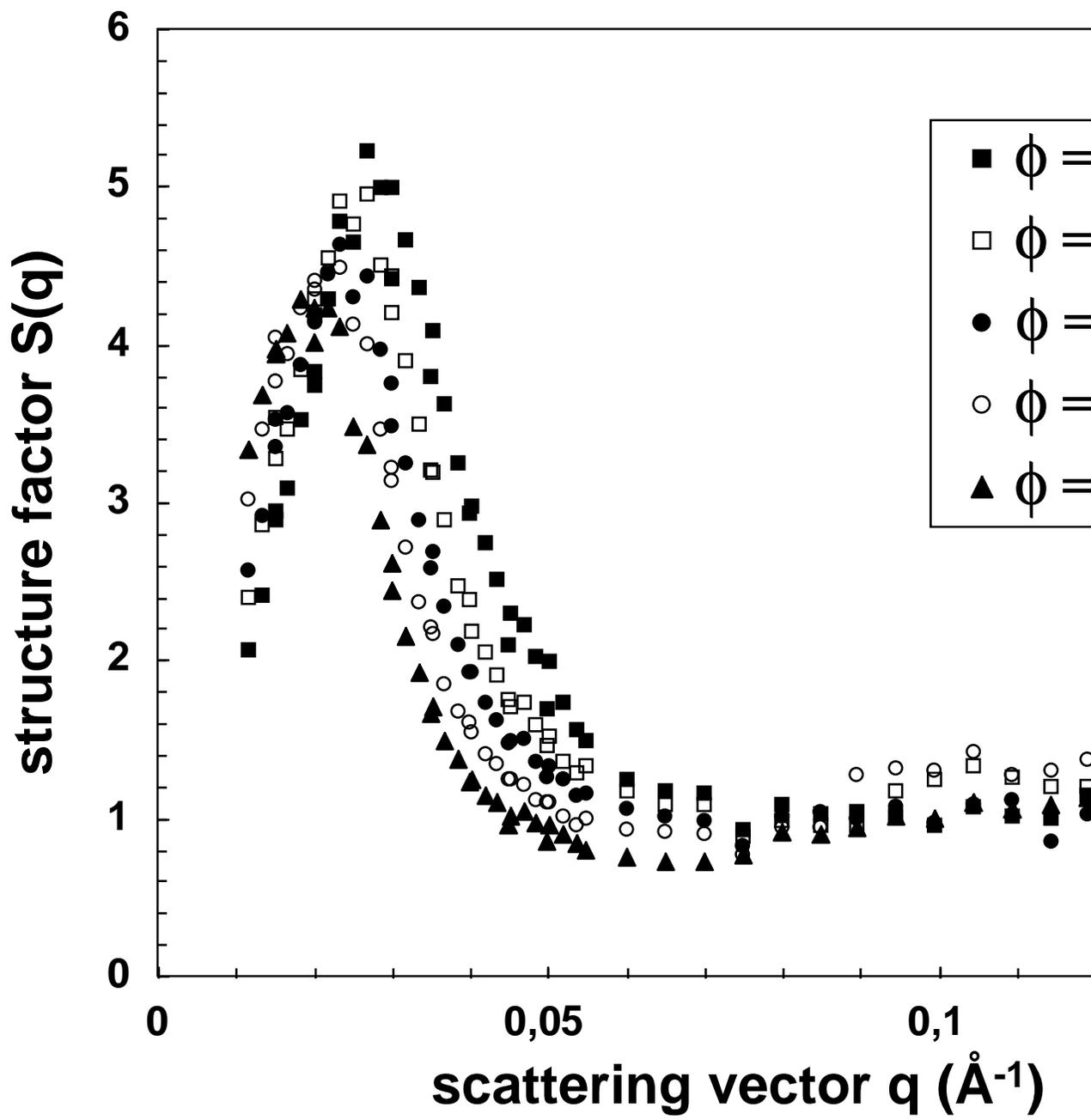
\end{document}